\documentclass[draftcls, onecolumn, journal, 12pt, letterpaper]{IEEEtran}

\usepackage[dvips]{graphicx}
\graphicspath{{../eps/}}
\DeclareGraphicsExtensions{.eps}
\usepackage{ifthen}
\newboolean{ONECOLUMN}
\setboolean{ONECOLUMN}{true}
\usepackage{amsfonts}
\usepackage{amssymb}
\usepackage{verbatim} 
\ifCLASSOPTIONcompsoc
	\usepackage[tight,normalsize,sf,SF]{subfigure}
\else
	\usepackage[tight,footnotesize]{subfigure}
\fi

\begin{document}

\title{Approximation of DAC Codeword Distribution for Equiprobable Binary Sources along Proper Decoding Paths}

\author{Yong~Fang
\thanks{This research was supported by NAFU Talent Fund Z111020901.}
\thanks{Y. Fang is with the College of Information Engineering, Northwest A\&F University, Shaanxi Yangling 712100, China (email: yfang79@gmail.com).}%
}

\maketitle

\begin{abstract}
Distributed Arithmetic Coding (DAC) is an effective implementation of Slepian-Wolf coding, especially for short data blocks. To research its properties, the concept of DAC codeword distribution along proper and wrong decoding paths has been introduced. For DAC codeword distribution of equiprobable binary sources along proper decoding paths, the problem was formatted as solving a system of functional equations. However, up to now, only one closed form was obtained at rate 0.5, while in general cases, to find the closed form of DAC codeword distribution still remains a very difficult task. This paper proposes three kinds of approximation methods for DAC codeword distribution of equiprobable binary sources along proper decoding paths: numeric approximation, polynomial approximation, and Gaussian approximation. Firstly, as a general approach, a numeric method is iterated to find the approximation to DAC codeword distribution. Secondly, at rates lower than 0.5, DAC codeword distribution can be well approximated by a polynomial. Thirdly, at very low rates, a Gaussian function centered at 0.5 is proved to be a good and simple approximation to DAC codeword distribution. A simple way to estimate the variance of Gaussian function is also proposed. Plenty of simulation results are given to verify theoretical analyses.
\end{abstract}


\section{Introduction}
\IEEEPARstart{C}{onsider} the problem of Slepian-Wolf Coding (SWC) with decoder Side Information (SI), i.e. the encoder compresses discrete source $X$ in the absence of $Y$, discretely-correlated SI. Slepian-Wolf theorem states that lossless compression is achievable at rates $R \geq H(X|Y)$ \cite{SlepianIT73}, where $H(X|Y)$ is the conditional entropy of $X$ given $Y$. Conventionally, channel codes, e.g., turbo codes \cite{FriasCL01} or Low-Density Parity-Check (LDPC) codes \cite{LiverisCL02}, are used to implement the SWC.

Ever since a long time ago, Arithmetic Coding (AC) has been proposed as the successor of Huffman coding to implement source coding and shows near-entropy performance \cite{RissanenIBMJRD76, RissanenIBMJRD79, RissanenActaPS79}. At the same time of high compression efficiency, the AC increases computational complexity and noise sensitivity of the bitstream. To reduce computational complexity, Quasi-Arithmetic Coding (QAC) has been introduced in \cite{HowardITC92}. To fight against noise sensitivity, redundancies are usually reinjected into the bitstream by different means. In \cite{ElmasryIEEPC99}, redundancies are reinjected into the bitstream in the form of parity-check bits. In \cite{SodagarICASSP00}, markers are inserted at known positions in the sequence of source symbols. In \cite{BoydTC97}, forbidden intervals are exploited for error detection. These approaches fall into the so-called Error Detecting AC (EDAC). The EDAC can be coupled with Automatic Repeat reQuest (ARQ) \cite{ElmasryTC99, ChouJSAC00, AnandTC01} or channel codes \cite{AnandTC01} to support error correction. To realize Error Correcting AC (ECAC), sequential decoding of arithmetic codes with forbidden intervals is proposed in \cite{PettijohnTC01}, whose complexity is reduced in \cite{DemirogluDCC01} by using Trellis-Coded Modulation (TCM) and List Viterbi decoding Algorithm (LVA). A soft decoding procedure is described in \cite{GuionnetTIP03}, whose counterpart for QAC appears in \cite{GuionnetEURASIPASP04}. The Maximum A Posteriori (MAP) decoding procedure is proposed and applied to image transmission \cite{GrangettoCL03, GrangettoTC05, GrangettoTIP06, GrangettoTIP07}.

Recently, the AC is also applied to implement the SWC. One approach is to allow overlapped intervals, which mirrors the work in \cite{BoydTC97}. Such examples include Distributed Arithmetic Coding (DAC) \cite{GrangettoCL07, GrangettoTSP09} and Overlapped Quasi-Arithmetic Coding (OQAC) \cite{ArtigasICIP07}. Another approach is to puncture some bits of AC bitstream, e.g. Punctured Quasi-Arithmetic Coding (PQAC) \cite{MalinowskiPCS09}, which mirrors the work in \cite{SodagarICASSP00}. There are also some variants of the DAC. The symmetric SWC is implemented by the time-shared DAC (TS-DAC) \cite{GrangettoMMSP07}. The rate-compatible DAC is proposed in \cite{GrangettoCL08}. Furthermore, decoder-driven adaptive DAC \cite{GrangettoICIP08} is proposed to estimate source probabilities on-the-fly.

We note that the ECAC and the DAC in fact generalize the classic AC in reverse directions. The ECAC encodes source $X$ at rates $R \geq H(X)/C \geq H(X)$ by introducing forbidden intervals, where $C$ is channel capacity. The forbidden intervals, corresponding to forbidden symbols, lead to a longer codeword due to narrowed final interval and also inject redundancies into the resulting bitstream. The decoder jointly exploits both the received bitstream and known channel parameters to reconstruct source $X$. The DAC encodes source $X$ at rates $H(X|Y) \leq R \leq H(X)$ by allowing overlapped intervals. The overlapped intervals, corresponding to ambiguous symbols, lead to a shorter codeword due to enlarged final interval and also induce ambiguities in the resulting bitstream. A soft joint decoder exploits both the received bitstream and $Y$to reconstruct $X$. 

Though it is well-known that the classic AC can achieve source entropy $H(X)$ theoretically, it is not clear whether the DAC can achieve conditional entropy $H(X|Y)$. If no, what is the performance limit of the DAC? Is it possible to improve its performance? If yes, how to realize it? Intuitively, before answering these questions, one may need to know how many branches will be generated during the DAC decoding. In addition, it may also be helpful to know the distribution of Hamming distances between decoding branches and source $X$.

As the first step, to analyze the properties of the DAC, \cite{FangSPL09} introduces the concept of codeword distribution, which seems promising for answering these questions. DAC codeword distribution is a function defined over interval $[0, 1)$. For equiprobable binary sources, both codeword distribution along proper decoding paths and codeword distribution along wrong decoding paths are researched. For codeword distribution along proper decoding paths, the problem is formatted as solving a system of functional equations including four constraints \cite{FangSPL09}. It is affirmed that rate $R = 0.5$ is a watershed: when $R>0.5$, DAC codeword distribution is an unsmooth function; while when $R\leq0.5$, DAC codeword distribution is a smooth function. Especially, a closed form is obtained at $R=0.5$. In spite of these achievements, it remains a very difficult task to find the closed form of codeword distribution along proper decoding paths in general. As for codeword distribution along wrong decoding paths, only some simulation results are reported in \cite{FangSPL09}, while problem formulation remains an open issue. It deserves to point out that the concept of codeword distribution can be easily extended to the ECAC.

This paper makes some advances on the work in \cite{FangSPL09}. Three approximation methods are proposed for codeword distribution of equiprobable binary sources along proper decoding paths: numeric approximation, polynomial approximation, and Gaussian approximation. Among them, numeric approximation is a general approach. At low rates ($R \leq 0.5$), polynomial approximation works well. To reduce computational complexity at very low rates, Gaussian approximation is used as an alternative to polynomial approximation.

This paper is arranged as follows. In Section \ref{sec:dac}, after a brief introduction to binary DAC codec, DAC decoding process is analyzed in detail to show the significance of DAC codeword distribution. Then the investigated problem is formulated in Section \ref{sec:format}. Section \ref{sec:numeric}, Section \ref{sec:polynomial}, and Section \ref{sec:gaussian} describe in detail numeric approximation, polynomial approximation, and Gaussian approximation, respectively, where simulation results are also reported. Finally, Section \ref{sec:conclusion} concludes this paper.

\section{Binary Distributed Arithmetic Coding}\label{sec:dac}
\subsection{Encoding}
Consider a binary source $X = \{x_i\}_{i=1}^{I}$ with bias probability $p = \textrm{Pr}(x_i = 1)$. In the classic AC, source symbol $x_i$ is iteratively mapped onto sub-intervals of $[0, 1)$, whose lengths are proportional to $(1-p)$ and $p$, giving rate $R = H(X)$. Instead, in the DAC \cite{GrangettoCL07, GrangettoTSP09}, sub-interval lengths are proportional to enlarged probabilities $(1-p)^\gamma$ and $p^\gamma$, where $H(X|Y)/H(X) \leq \gamma \leq 1$, giving rate $R = \gamma H(X) \geq H(X|Y)$. For conciseness, we refer to $\gamma$ as overlap coefficient hereinafter. More specifically, symbols $x_i = 0$ and $x_i = 1$ correspond to sub-intervals $[0, (1-p)^\gamma)$ and $[1-p^\gamma, 1)$, respectively. It means that to fit the $[0, 1)$ interval, the sub-intervals have to be partially overlapped. This overlapping leads to a larger final interval, and hence a shorter codeword. However, as a cost, the decoder can not decode $X$ unambiguously without $Y$.

Note that when $\gamma \geq 1/C \geq 1$, where $C$ is channel capacity, it becomes the ECAC.

\subsection{Decoding}
To describe the decoding process, a ternary symbol set $\{0, \mathcal{A}, 1\}$ is defined, where $\mathcal{A}$ represents the ambiguous symbol. Let $C_X$ be DAC codeword and $\tilde{x}_i$ be the $i$-th decoded symbol, then
\begin{equation}
\setlength{\nulldelimiterspace}{0pt}
\tilde{x}_i = \left\{
\begin{IEEEeqnarraybox} [\relax] [c] {l's}
0, &$0 \leq C_X < 1 - p^\gamma$ \\
\mathcal{A}, &$1 - p^\gamma \leq C_X < (1-p)^\gamma$\\
1, &$(1-p)^\gamma \leq C_X < 1$%
\end{IEEEeqnarraybox}.
\right.
\end{equation}
After $\tilde{x}_i$ is decoded, if $\tilde{x}_i = \mathcal{A}$, the decoder will perform a branching: two candidate branches are generated, corresponding to two alternative symbols $x_i = 0$ and $x_i = 1$. For each new branch, its metric is updated and the corresponding interval is selected for next iteration. To reduce complexity, every time a symbol is decoded, the decoder uses the $M$-algorithm to keep at most $M$ paths with the best partial metric, and prunes others \cite{GrangettoCL07, GrangettoTSP09}. Finally, after all source symbols are decoded, the path with the best metric is output as the estimate of $X$. As for detailed performance comparisons between DAC and LDPC-based SWC, please refer to \cite{GrangettoTSP09}.

\subsection{Discussion}
It deserves to point out that during DAC decoding, the metric of each path is indeed the Hamming distance between this path and SI $Y$. As we know, each DAC codeword defines a set of possible decoding paths and each possible decoding path corresponds to a sequence of decoded symbols. However, among all possible decoding paths, there is one and only one proper path which corresponds to source $X$. Let $\tilde{X} = \{\tilde{x}_i\}_{i=1}^{I}$ be a sequence of decoded symbols. Let $D(Y, \tilde{X})$ be the Hamming distance between $Y$ and $\tilde{X}$. Similarly, $D(X, \tilde{X})$ and $D(X, Y)$ are also defined. Obviously,
\begin{equation}
D(Y, \tilde{X}) \leq D(X, Y) + D(X, \tilde{X}).
\end{equation}
The task of a DAC decoder is in fact to find a path $X'$ that minimizes $D(Y, \tilde{X})$, i.e.
\begin{equation}
X' = \arg\min_{\tilde{X}} {D(Y, \tilde{X})}.
\end{equation}
However, this is not always followed by $D(X, X') = 0$. If $D(X, X') \neq 0$, then a decoding failure occurs. To find the probability of decoding failure, we need to know the distribution of $D(Y, \tilde{X})$ and $D(X, \tilde{X})$. 

Though it is very difficult to find the distribution of $D(Y, \tilde{X})$ and $D(X, \tilde{X})$, this problem can be tackled by means of DAC codeword distribution. As shown in \cite{FangSPL09}, if we know codeword distributions along proper and wrong decoding paths, it seems promising to find the number of possible decoding paths and the distribution of $D(Y, \tilde{X})$ and $D(X, \tilde{X})$. 

The rest of this paper makes some advances on DAC codeword distribution along proper decoding paths.

\section{Problem Formulation}\label{sec:format}
To simplify the analysis, we consider an infinite-length, stationary, and equiprobable binary source $X = \{x_i\}_{i=1}^{\infty}$. As $p = 0.5$, symbols $x_i = 0$ and $x_i = 1$ correspond to sub-intervals $[0, q)$ and $[1-q, 1)$ respectively, where $q = 0.5^{\gamma}$. The resulting rate $R = \gamma H(X) = \gamma$.

\begin{figure}
\centering
\ifCLASSOPTIONtwocolumn
	\includegraphics[width=\linewidth]{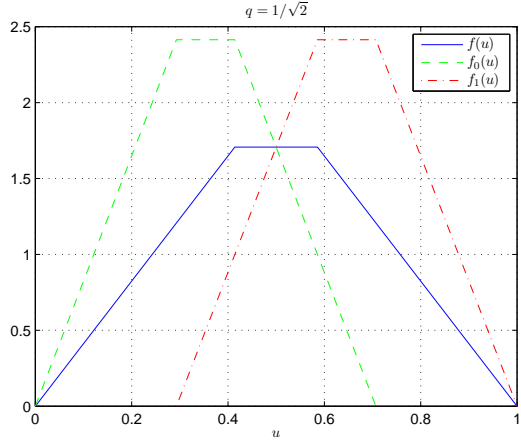}
\else
	\includegraphics[width=.5\linewidth]{illustration.eps}
\fi
\caption{Illustrations of $f(u)$, $f_0(u)$, and $f_1(u)$ for $q=1/\sqrt{2}$. $f(u)$ is symmetric around $u=0.5$, i.e. $f(u) = f(1-u)$. $f(u)$, $f_0(u)$, and $f_1(u)$ have the same shape. $f(u) = (f_0(u) + f_1(u))/2$. $f_0(u)$ can be obtained by first squeezing $f(u)$ by $q$ times along $x$-axis and then streching $f(u)$ by $1/q$ times along $y$-axis, i.e. $f_0(u) = f(u/q)/q$. $f_1(u)$ can be obtained by shifting $f_0(u)$ right by $(1-q)$, i.e. $f_1(u) = f_0(u-(1-q))$. Due to the symmetry, $f_1(u) = f_0(1-u)$. $f(u)$, $f_0(u)$, and $f_1(u)$ intersect at $u = 0.5$, i.e. $f(0.5) = f_0(0.5) = f_1(0.5)$. Hence $qf(0.5) = f(0.5/q)$.}
\label{fig:dacdis}
\end{figure}

Let $C_X$ be the DAC codeword of $X$ and $f(u)$ $(0 \leq u < 1)$ be the distribution of $C_X$, then 
\begin{equation}
\int_0^1 {f(u)du} = 1. 
\end{equation}
Due to the symmetry, we have
\begin{equation}
f(u) = f(1-u), \quad 0 < u < 1.
\end{equation}
Symbols $x_1 = 0$ and $x_1=1$ correspond to intervals $[0, q)$ and $[1-q, 1)$, respectively. If $x_1=0$, the remaining sequence $X_2=\{x_i\}_{i=2}^{\infty}$ will be iteratively mapped onto the sub-intervals of $[0, q)$; otherwise, $X_2$ will be iteratively mapped onto the sub-intervals of $[1-q, 1)$. Let $C_{X_2}^0$ be the DAC codeword of $X_2$ given $x_1=0$ and $f_0(u)$ be the distribution of $C_{X_2}^0$, then 
\begin{equation}
\int_0^q {f_0(u)du} = 1. 
\end{equation}
Since $X$ is infinite-length and stationary, $f_0(u)$ must have the same shape as $f(u)$, i.e.,  
\begin{equation}
f_0(u) = f(u/q)/q, \quad 0 \leq u < q.
\end{equation}
Similarly, let $C_{X_2}^1$ be the DAC codeword of $X_2$ given $x_1=1$ and $f_1(u)$ be the distribution of $C_{X_2}^1$, then 
\ifCLASSOPTIONtwocolumn
	\begin{eqnarray}
		f_1(u) = f_0(u-(1-q)) = f(\frac{u-(1-q)}{q})/q,\nonumber\\
		\qquad\qquad\qquad\qquad\qquad (1-q) \leq u < 1.%
	\end{eqnarray}
\else
	\begin{equation}
		f_1(u) = f_0(u-(1-q)) = f(\frac{u-(1-q)}{q})/q, \quad (1-q) \leq u < 1.
	\end{equation}
\fi
Due to the symmetry,
\begin{equation}
f_1(u) = f_0(1-u).
\end{equation}
The relations between $f(u)$, $f_0(u)$, and $f_1(u)$ can be illustrated by Fig.\ref{fig:dacdis}. Obviously, 
\ifCLASSOPTIONtwocolumn
	\begin{eqnarray}
		f(u) 	&=& \mathrm{Pr}(x_1=0)f_0(u) + \mathrm{Pr}(x_1=1)f_1(u)\nonumber\\ 
					&=& (f_0(u) + f_1(u))/2.%
	\end{eqnarray}
\else
	\begin{equation}
		f(u) = \mathrm{Pr}(x_1=0)f_0(u) + \mathrm{Pr}(x_1=1)f_1(u) = (f_0(u) + f_1(u))/2.%
	\end{equation}
\fi
Hence, $f(u)$, $f_0(u)$, and $f_1(u)$ intersect at $u = 0.5$, i.e. $f(0.5) = f_0(0.5) = f_1(0.5)$. Thus 
\begin{equation}
qf(0.5) = f(0.5/q).
\end{equation}

\subsection{Classic AC}
When $q = 0.5$, it is just the classic AC. Then
\begin{equation}
\setlength{\nulldelimiterspace}{0pt}
f(u) = \left\{
\begin{IEEEeqnarraybox} [\relax] [c] {l's}
f(2u), &$0 \leq u < 0.5$ \\
f(2u-1), &$0.5 \leq u < 1$%
\end{IEEEeqnarraybox}.
\right.
\end{equation}
It is easy to prove $f(u) \equiv 1$ ($0 \leq u < 1$). This is a uniform distribution, so the classic AC can achieve source entropy theoretically.

\subsection{Distributed AC}
When $0.5 < q <1$, sub-intervals $[0, q)$ and $[1-q, 1)$ are partially overlapped, so $f(u)$ is a piecewise-defined function.
\subsubsection{$0 \leq u < (1-q)$} It this interval, $f_1(u) = 0$, so 
\begin{equation}
f(u) = f_0(u)/2 = f(u/q)/(2q). 
\end{equation}
Since $f(0) = f(0/q)/(2q)$, we have $f(0) = 0$. 
\subsubsection{$q \leq u < 1$} In this interval, $f_0(u) = 0$, so 
\begin{equation}
f(u) = f_1(u)/2 = f(\frac{u-(1-q)}{q})/(2q). 
\end{equation}
\subsubsection{$1-q \leq u < q$} In this interval, we have  
\begin{equation}
f(u) = \frac{f(\frac{u}{q}) + f(\frac{u-(1-q)}{q})}{2q}. 
\end{equation}

\subsection{A Closed Form of $f(u)$ at $q=1/\sqrt{2}$}
Generally, it is very difficult to obtain the closed form of $f(u)$. In \cite{FangSPL09}, only one closed form is obtained at $q = 1/\sqrt{2}$ (i.e. $\gamma = 0.5$): 
\begin{equation}
f(u) = \left\{
\begin{IEEEeqnarraybox} [\relax] [c] {l's}
\frac{u}{3\sqrt{2}-4}, &$0 \leq u \leq \sqrt{2}-1$\\
\frac{1}{2-\sqrt{2}}, &$\sqrt{2}-1 \leq u \leq 2-\sqrt{2}$\\
\frac{1-u}{3\sqrt{2}-4}, &$2-\sqrt{2} \leq u \leq 1$%
\end{IEEEeqnarraybox}.
\right.
\end{equation}

\subsection{Zeros of $f(u)$ at High Rates}
It is proved in \cite{FangSPL09} that when $0.5 < q \leq \frac{\sqrt{5}-1}{2}$ (corresponds to $0.6942 \leq \gamma < 1$), $f(\frac{q^n}{q+1}) = f(1-\frac{q^n}{q+1}) = 0$, $\forall n \in \mathbb{N}$.

\section{Numeric Approximation}\label{sec:numeric}
Though a special closed form of $f(u)$ is found for $p = \gamma = 0.5$ in \cite{FangSPL09}, the procedure is very complex. In general, the closed form of $f(u)$ does not exist. As a universal approach, we propose a numeric method for finding $f(u)$. This method is described in detail below.

\subsection{Discretization}
We divide the interval $[0, 1]$ into $N$ uniform cells. Let $\Delta = 1/N$. Then $f(u)$ can be approximated by $f(n\Delta)$, where $n \in \mathcal{I}_N = \{0,1,...,N\}$, given a large $N$.

\subsection{Initialization}
Let $f^{(t)}(n\Delta)$ be the estimate of $f(n\Delta)$ after $t$ iterations. Before iteration, $f^{(0)}(n\Delta)$ need to be initialized. Though arbitrary initialization is allowed, we recommend uniform initialization, i.e. $f^{(0)}(n\Delta) \equiv 1$, where $n \in \mathcal{I}_N$.

\subsection{Iteration}
Let $L = \lfloor N(1-q) \rfloor = N - \lceil Nq \rceil$ and $H = \lceil Nq \rceil$. Then the iteration is run as follows.

\subsubsection{$0 \leq n \leq L$} This corresponds to interval $0 \leq u \leq (1-q)$, hence 
\begin{equation}
f^{(t)}(n\Delta) = \frac{f^{(t-1)}(h_{0,N}(n/q)\Delta)}{2q},
\end{equation}
where 
\begin{equation}
h_{a,b}(x) = \left\{
\begin{IEEEeqnarraybox} [\relax] [c] {l's}
a, &$round(x) < a < b$ \\
round(x), &$a \leq round(x) \leq b$\\
b, &$a < b < round(x)$%
\end{IEEEeqnarraybox}.
\right.
\end{equation}

\subsubsection{$H \leq n \leq N$} This corresponds to interval $q \leq u \leq 1$. Because $L+H = N$,
\begin{equation}
f^{(t)}(n\Delta) = f^{(t)}((N-n)\Delta). 
\end{equation}

\subsubsection{$L < n < H$} This corresponds to interval $(1-q) < u < q$, hence  
\begin{equation}
f^{(t)}(n\Delta) = \frac{f^{(t-1)}(h_{0,N}(n/q)\Delta) + f^{(t-1)}(h_{0,N}(\frac{n-L}{q})\Delta)}{2q}. 
\end{equation}

\subsection{Normalization}
Recall the constraint $\int_{0}^{1}{f(u)du} = 1$, we have 
\begin{equation}
\sum_{n=0}^{N}{f^{(t)}(n\Delta)\Delta} = 1, 
\end{equation}
i.e.
\begin{equation}
\sum_{n=0}^{N}{f^{(t)}(n\Delta)} = 1/\Delta = N. 
\end{equation}
Let $\sum_{n=0}^{N}{f^{(t)}(n\Delta)} = \Omega$, then $f^{(t)}(n\Delta)$ should be normalized as below:
\begin{equation}
f^{(t)}(n\Delta) = \frac{Nf^{(t)}(n\Delta)}{\Omega}. 
\end{equation}

\subsection{Termination}
We use the Mean Squared Error (MSE) between two successive iterations as a measurement to terminate the iteration. Let $\delta$ be a small quantity. When
\begin{equation}
\mathrm{MSE}^{(t)} = \frac{1}{N+1}\sum_{n=0}^{N}{(f^{(t)}(n\Delta) - f^{(t-1)}(n\Delta))^2} < \delta,
\end{equation}
the iteration is terminated.

\subsection{Simulation Results}
\begin{figure*}
\centering
\subfigure[]{\includegraphics[width=.5\linewidth]{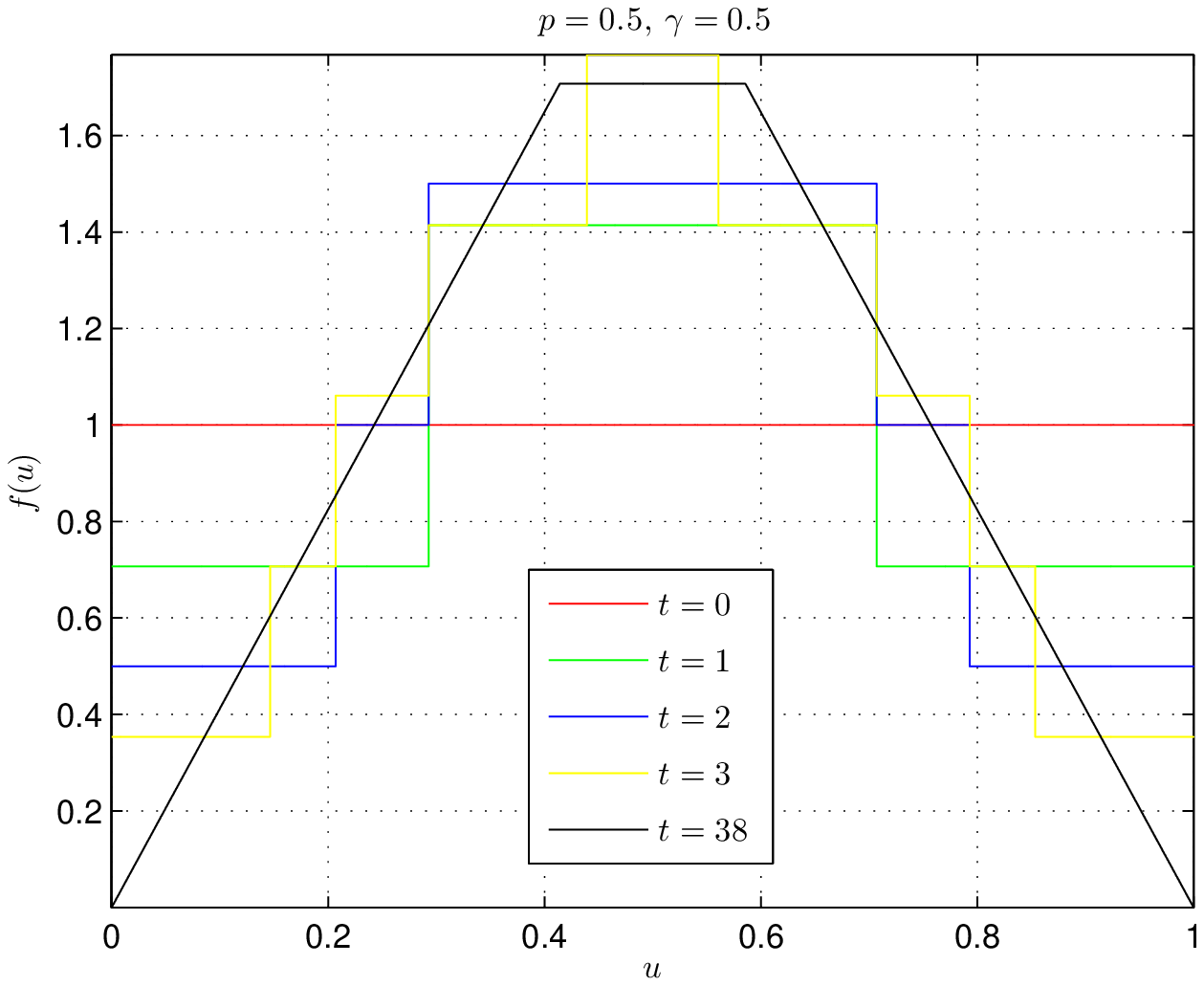}}\subfigure[]{\includegraphics[width=.5\linewidth]{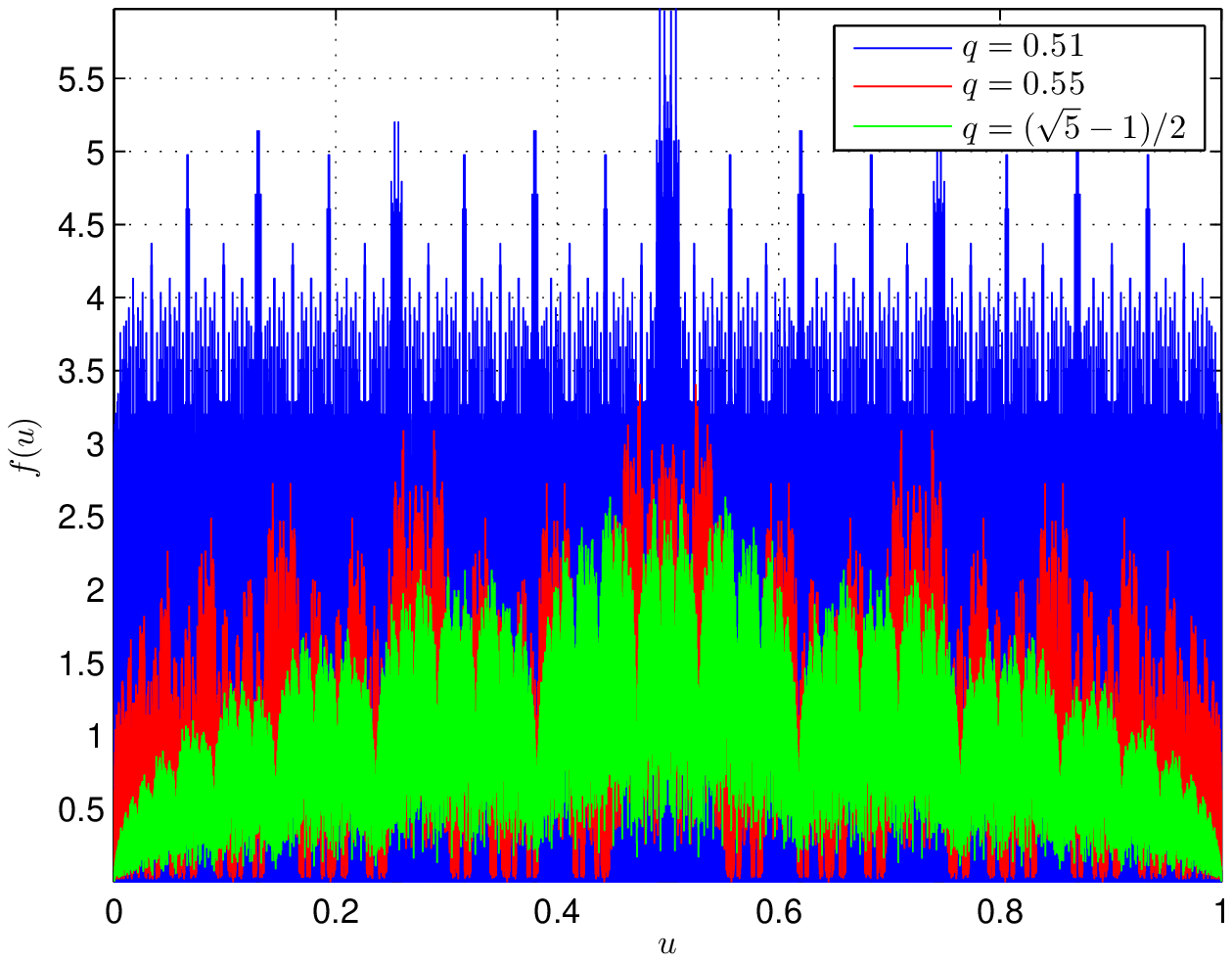}}
\subfigure[]{\includegraphics[width=.5\linewidth]{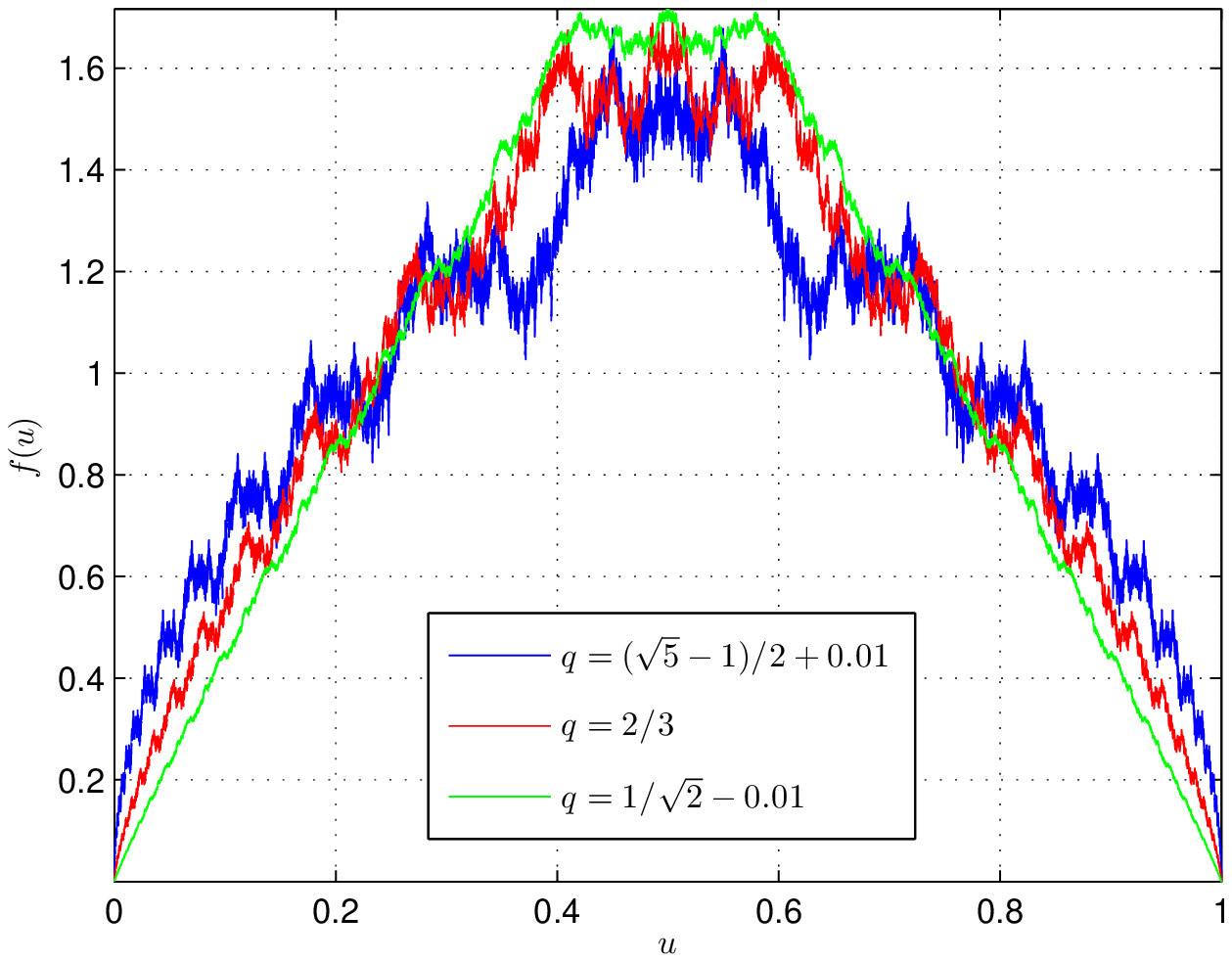}}\subfigure[]{\includegraphics[width=.5\linewidth]{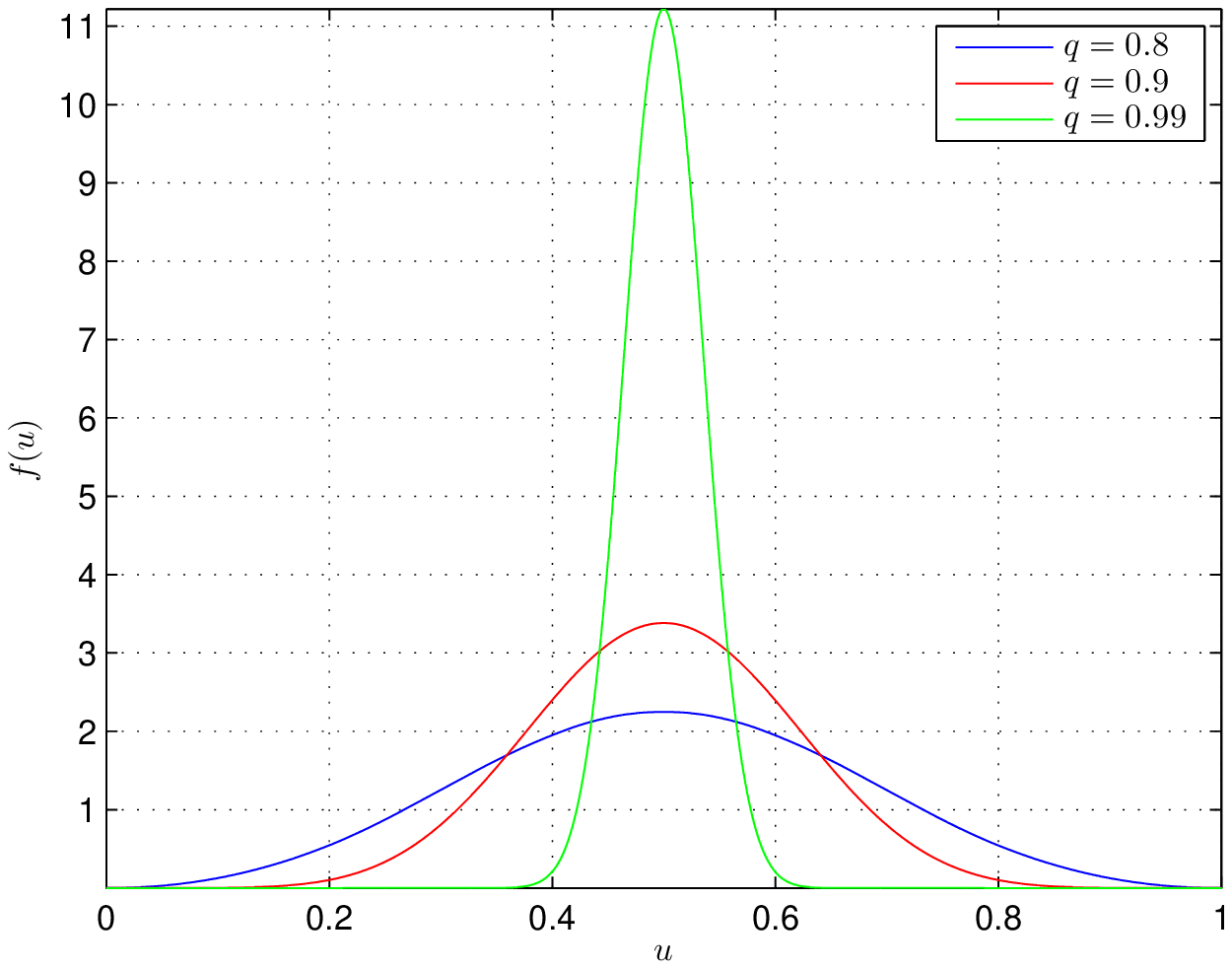}}
\caption{Simulation curves of numeric approximation to $f(u)$, where $N = 10^5$. All these results coincides with those given in \cite{FangSPL09}, meaning that numeric approximation is well justified. (a) Evolution of $f^{(t)}(n\Delta)$ with respect to $t$ for $q$ = $1/\sqrt{2}$. As $t$ increases, $f^{(t)}(n\Delta)$ converges to $f(u)$. $\mathrm{MSE}^{(38)}<10^{-10}$. (b) Some results for $q \in (0.5, (\sqrt{5}-1)/2]$. $\mathrm{MSE}^{(586)}<10^{-4}$ for $q=0.51$. $\mathrm{MSE}^{(70)}<10^{-4}$ for $q=0.55$. $\mathrm{MSE}^{(51)}<10^{-4}$ for $q=(\sqrt{5}-1)/2$. (c) Some results for $q \in ((\sqrt{5}-1)/2, 1/\sqrt{2})$. $\mathrm{MSE}^{(85)}<10^{-9}$ for $q=(\sqrt{5}-1)/2+0.01$. $\mathrm{MSE}^{(63)}<10^{-9}$ for $q=2/3$. $\mathrm{MSE}^{(52)}<10^{-9}$ for $q=1/\sqrt{2}-0.01$. (d) Some results for $q \in [1/\sqrt{2}, 1)$. $\mathrm{MSE}^{(39)}<10^{-10}$ for $q=0.8$. $\mathrm{MSE}^{(54)}<10^{-10}$ for $q=0.9$. $\mathrm{MSE}^{(540)}<10^{-9}$ for $q=0.99$.}
\label{fig:numeric}
\end{figure*}

Fig. \ref{fig:numeric} includes some results regarding numeric approximation. All results reported in Fig. \ref{fig:numeric} are obtained with $N=10^5$. 

To show how $f^{(t)}(n\Delta)$ converges to $f(u)$, the evolution of $f^{(t)}(n\Delta)$ with $t$ is plotted in Fig. \ref{fig:numeric}(a). We find that after 38 iterations, successive MSE has been less than $10^{-10}$.

It was affirmed in \cite{FangSPL09} that $(\sqrt{5}-1)/2$ and $1/\sqrt{2}$ are two watersheds that divide interval $(0.5, 1)$ of $q$ into three sub-intervals: $(0.5, (\sqrt{5}-1)/2]$, $((\sqrt{5}-1)/2, 1/\sqrt{2})$, and $[1/\sqrt{2}, 1)$, because $f(u)$ shows very different properties in these three sub-intervals. As in \cite{FangSPL09}, for each sub-interval of $q$, some simulation results are reported in Figs. \ref{fig:numeric}(b)-(d). All these results coincide with those given in \cite{FangSPL09} perfectly. Fig. \ref{fig:numeric}(b) confirms the zeros of $f(u)$ at high rates. Fig. \ref{fig:numeric}(d) shows that $f(u)$ becomes smooth at low rates.

In different sub-intervals, numeric approximation shows very different simulation precision and computational complexity. Firstly, we consider simulation precision. For $q \in (0.5, (\sqrt{5}-1)/2]$, tens of iterations are needed to make successive MSE less than $10^{-4}$, while for $q \in ((\sqrt{5}-1)/2, 1/\sqrt{2})$, tens of iterations have made successive MSE less than $10^{-9}$. For $q \in [1/\sqrt{2}, 1)$, tens of iterations can even make successive MSE less than $10^{-10}$. Secondly, we consider computational complexity. We find that $q=1/\sqrt{2}$ needs the fewest iterations. As $q$ departs from $1/\sqrt{2}$ (increase or decrease), computational complexity increases, i.e. more iterations are needed to reach the same successive MSE. Thirdly, as $q$ approaches to 0.5 or 1, simulation precision is sharply degraded, or in other words, computational complexity increases sharply. For example, when $q=0.51$, 586 iterations are needed to make successive MSE less than $10^{-4}$, while for other $q$ in the same sub-intervals (e.g. 0.55 and $(\sqrt{5}-1)/2$), tens of iterations are enough to reach the same precision. Similar phenomenon is also observed for $q=0.99$.

It may be an interesting issue to improve simulation precision and accelerate convergence speed of $f(u)$, especially for $q$ close to 0.5 or 1.

\section{Polynomial Approximation at Low Rates}\label{sec:polynomial}
It was affirmed in \cite{FangSPL09} that $f(u)$ is a smooth function when $q \geq 1/\sqrt{2}$, i.e. $R \leq 0.5$. This property suggests that polynomials may be good approximation to $f(u)$ at low rates ($R \leq 0.5$). Below we propose polynomial approximation to $f(u)$ for $1/\sqrt{2} \leq q < 1$.

To simplify the analysis, we exploit the symmetry and consider only the left half of $f(u)$
\begin{equation}
f(u) = \left\{
\begin{IEEEeqnarraybox} [\relax] [c] {l's}
f(u/q)/(2q), &$0 \leq u \leq (1-q)$\\
\frac{f(\frac{u}{q}) + f(\frac{u-(1-q)}{q})}{2q}, &$(1-q) \leq u \leq 0.5$%
\end{IEEEeqnarraybox}.
\right.
\label{eq:origin}
\end{equation}
We rewrite (\ref{eq:origin}) as 
\begin{equation}
f(u) = \left\{
\begin{IEEEeqnarraybox} [\relax] [c] {l's}
2qf(qu), & $0 \leq u \leq v_1$\\
2qf(qu) - f(u - v_1), & $v_1 \leq u \leq 0.5$%
\end{IEEEeqnarraybox}.
\right.
\end{equation}
where $v_n = (1-q)/q^n$, $n \in \mathbb{N}$. Note that $v_1 < 0.5$ when $q \geq 1/\sqrt{2}$. Hence, $f(u)$ is a piecewise-defined function over interval $[0, 0.5]$. 

At first, in sub-interval $[0, v_1]$, $f(u)$ can be obtained by solving functional equation $f(u)=2qf(qu)$ \cite{FuncEq}
\begin{equation}
f(u) = \phi(u) = \Theta(u) u^\lambda, \quad 0 \leq u \leq v_1,
\label{eq:phi}
\end{equation}
where $\lambda = (1-\gamma)/\gamma$ and $\Theta(u) = \Theta(uq^k)$, $\forall k \in \mathbb{Z}$. 

Then, we need to determine $f(u)$ in sub-interval $[v_1, 0.5]$. Because $qu < u$ and $u - v_1 < u$, it is possible to recursively map sub-interval $[v_1, 0.5]$ onto sub-interval $[0, v_1]$ by scaling down or shifting $u$, over which $f(u)$ has been given by (\ref{eq:phi}), i.e. 
\begin{equation}
\left\{
\begin{IEEEeqnarraybox} [\relax] [c] {l's}
f(qu) = \phi(qu), & $v_1 \leq u \leq v_2$\\
f(u-v_1) = \phi(u-v_1), & $v_1 \leq u \leq 2v_1$%
\end{IEEEeqnarraybox}.
\right.
\end{equation}
This is the key to solving this problem.

It is easy to prove that $u - v_1 < qu$ for $u \in [v_1, 0.5]$. Hence,
\begin{equation}
f(u) = 2qf(qu) - \phi(u-v_1), \quad v_1 \leq u \leq 2v_1.
\end{equation}
On solving $2v_1 = 0.5$, we obtain $q=0.8$. Hereinafter, to facilitate our description, we divide interval $1/\sqrt{2} \leq q < 1$ into two sub-intervals $1/\sqrt{2} \leq q \leq 0.8$ (corresponding to $0.5 \leq 2v_1$) and $0.8 < q < 1$ (corresponding to $2v_1 < 0.5$).

\subsection{$1/\sqrt{2} \leq q \leq 0.8$}\label{subsec:polyA}
In this sub-interval, since $0.5 \leq 2v_1$, we have
\begin{equation}
f(u) = \left\{
\begin{IEEEeqnarraybox} [\relax] [c] {l's}
\phi(u), &$0 \leq u \leq v_1$\\
2qf(qu) - \phi(u-v_1), &$v_1 \leq u \leq 0.5$%
\end{IEEEeqnarraybox}.
\right.
\end{equation}
Hence, we need to consider only the term $2qf(qu)$. Depending on the relations between $v_n$ and 0.5, this sub-interval can be further divided into three smaller sub-intervals.

\subsubsection{$0.5 \leq v_2$} On solving $v_2 = 0.5$, we obtain $q = \sqrt{3}-1$, so this sub-interval corresponds to $1/\sqrt{2} \leq q \leq \sqrt{3}-1$. Since $0.5 \leq v_2$, we have $qu \leq v_1$ for $u \in [v_1, 0.5]$, i.e. $f(qu) = \phi(qu)$. Remember $\phi(u) \equiv 2q\phi(qu)$. Thus
\begin{equation}
f(u) = \left\{
\begin{IEEEeqnarraybox} [\relax] [c] {l's}
\phi(u), &$0 \leq u \leq v_1$\\
\phi(u) - \phi(u-v_1), &$v_1 \leq u \leq 0.5$%
\end{IEEEeqnarraybox}.
\right.
\end{equation}
As affirmed in \cite{FangSPL09}, $f(u)$ is a smooth function for $q \geq 1/\sqrt{2}$. Hence we approximate $\Theta(u)$ by a const $c$ and then obtain
\begin{equation}
f(u) \approx \left\{
\begin{IEEEeqnarraybox} [\relax] [c] {l's}
cu^\lambda, &$0 \leq u \leq v_1$\\
cu^\lambda - c(u-v_1)^\lambda, &$v_1 \leq u \leq 0.5$%
\end{IEEEeqnarraybox}.
\right.
\end{equation}
Now we need to determine $c$. Let us integrate $f(u)$ over interval [$0, 0.5$]
\begin{eqnarray} 
\int_{0}^{0.5}{f(u)du} 	
											 	&=& c\left(\int_{0}^{0.5}{u^\lambda du} - \int_{v_1}^{0.5}{(u-v_1)^\lambda du}\right)\nonumber\\
											 	&=& \frac{c(u^{\lambda+1}|_{0}^{0.5} - (u-v_1)^{\lambda+1}|_{v_1}^{0.5})}{\lambda + 1}\nonumber\\
											 	&=& \frac{c(0.5^{\lambda+1} - (0.5-v_1)^{\lambda+1})}{\lambda + 1} = 0.5.%
\end{eqnarray}
Thus,
\begin{equation}
c = \frac{0.5(\lambda + 1)}{0.5^{\lambda+1} - (0.5-v_1)^{\lambda+1}}.
\end{equation}
Due to $\lambda+1 = 1/\gamma$,
\begin{equation}
c = \frac{1}{2\gamma(0.5^{(1/\gamma)} - (0.5-v_1)^{(1/\gamma)})}.
\end{equation}

\subsubsection{$v_2 < 0.5 \leq v_3$} On solving $v_3 = 0.5$, we obtain $q \approx 0.77$, so this sub-interval corresponds to $\sqrt{3}-1 < q \leq 0.77$. At first, it can be obtained directly
\begin{equation}
f(u) = \left\{
\begin{IEEEeqnarraybox} [\relax] [c] {l's}
\phi(u), &$0 \leq u \leq v_1$\\
\phi(u) - \phi(u-v_1), &$v_1 \leq u \leq v_2$%
\end{IEEEeqnarraybox}.
\right.
\end{equation}
Then, for $u \in [v_2, 0.5]$, we have $qu \in [v_1, v_2]$, i.e. $f(qu) = \phi(qu) - \phi(qu-v_1)$. Thus
\ifCLASSOPTIONtwocolumn
	\begin{eqnarray}
		f(u) 	&=& 2qf(qu) - \phi(u - v_1)\nonumber\\
					&=& 2q(\phi(qu) - \phi(qu-v_1)) - \phi(u - v_1),\nonumber\\
					& & v_2 \leq u \leq 0.5.%
	\end{eqnarray}
\else
	\begin{eqnarray}
		f(u) 	&=& 2qf(qu) - \phi(u - v_1)\nonumber\\
					&=& 2q(\phi(qu) - \phi(qu-v_1)) - \phi(u - v_1),\quad v_2 \leq u \leq 0.5.%
	\end{eqnarray}
\fi			
Because $2q\phi(qu-v_1) = 2q\phi(q(u-v_2)) = \phi(u-v_2)$, we obtain
\begin{equation}
f(u) = \phi(u) - \sum_{i=1}^{2}{\phi(u-v_i)}, \quad v_2 \leq u \leq 0.5.%
\end{equation}
Therefore, we can obtain the following approximation
\begin{equation}
	f(u) \approx \left\{
		\begin{IEEEeqnarraybox} [\relax] [c] {l's}
			cu^\lambda, &$0 \leq u \leq v_1$\\
			cu^\lambda - c(u-v_1)^\lambda, &$v_1 \leq u \leq v_2$\\
			cu^\lambda - c\sum_{i=1}^{2}{(u-v_i)^\lambda}, &$v_2 \leq u \leq 0.5$%
		\end{IEEEeqnarraybox},
	\right.
\end{equation}
where
\begin{equation}
c = \frac{1}{2\gamma(0.5^{(1/\gamma)} - \sum_{i=1}^{2}{(0.5-v_i)^{(1/\gamma)}})}.
\end{equation}

\subsubsection{$v_3 < 0.5 \leq v_4$}On solving $v_4 = 0.5$, we obtain $q \approx 0.8$, so this sub-interval corresponds to $0.77 < q \leq 0.8$. By iterations, we can obtain 
\begin{equation}
f(u) \approx \left\{
\begin{IEEEeqnarraybox} [\relax] [c] {l's}
cu^\lambda, &$0 \leq u \leq v_1$\\
cu^\lambda - c(u-v_1)^\lambda, &$v_1 \leq u \leq v_2$\\
cu^\lambda - c\sum_{i=1}^{2}{(u-v_i)^\lambda}, &$v_2 \leq u \leq v_3$\\
cu^\lambda - c\sum_{i=1}^{3}{(u-v_i)^\lambda}, &$v_3 \leq u \leq 0.5$%
\end{IEEEeqnarraybox},
\right.
\end{equation}
where
\begin{equation}
c = \frac{1}{2\gamma(0.5^{(1/\gamma)} - \sum_{i=1}^{3}{(0.5-v_i)^{(1/\gamma)}})}.
\end{equation}

\subsection{$0.8 < q < 1$}
The problem becomes very complex in this sub-interval because $f(u-v_1) = \phi(u-v_1)$ does not hold for $u \in [2v_1, 0.5]$ so that we need to deal with not only $2qf(qu)$ but also $f(u-v_1)$.

Let us consider a simple case first, i.e. $v_1 < 0.5-v_1 \leq v_2$, which corresponds to sub-interval $0.8 < q \leq \sqrt{2/3}$. We have $u-v_1 \in [v_1, v_2]$ for $u \in [2v_1, 0.5]$. Hence
\begin{equation}
f(u-v_1) = \phi(u - v_1) - \phi(u - 2v_1), \quad 2v_1 \leq u \leq 0.5. 
\end{equation}
Therefore, the problem becomes
\ifCLASSOPTIONtwocolumn
	\begin{equation}
		f(u) = \left\{
			\begin{IEEEeqnarraybox} [\relax] [c] {l's}
				2qf(qu), 	\qquad\qquad\qquad\qquad\, 0 \leq u \leq v_1\\
				2qf(qu) - \phi(u - v_1), \qquad v_1 \leq u \leq 2v_1\\
				2qf(qu) - (\phi(u - v_1) - \phi(u - 2v_1)),&\\
				\qquad\qquad\qquad\qquad\qquad\quad 2v_1 \leq u \leq 0.5%
			\end{IEEEeqnarraybox}.
		\right.
	\end{equation}
\else
	\begin{equation}
		f(u) = \left\{
			\begin{IEEEeqnarraybox} [\relax] [c] {l's}
				2qf(qu), & $0 \leq u \leq v_1$\\
				2qf(qu) - \phi(u - v_1), & $v_1 \leq u \leq 2v_1$\\
				2qf(qu) - (\phi(u - v_1) - \phi(u - 2v_1)), & $2v_1 \leq u \leq 0.5$%
			\end{IEEEeqnarraybox}.
		\right.
	\end{equation}
\fi
Now we need to deal with only $2qf(qu)$, which has been discussed in detail in Section \ref{subsec:polyA}.

For $\sqrt{2/3} < q < 1$, the idea is the same but the procedure becomes more and more complicated as $q$ increases. Therefore, at very low rates, polynomial approximation is not a good choice.

\subsection{Simulation Results}
\begin{figure*}
\centering
\subfigure[]{\includegraphics[width=.5\linewidth]{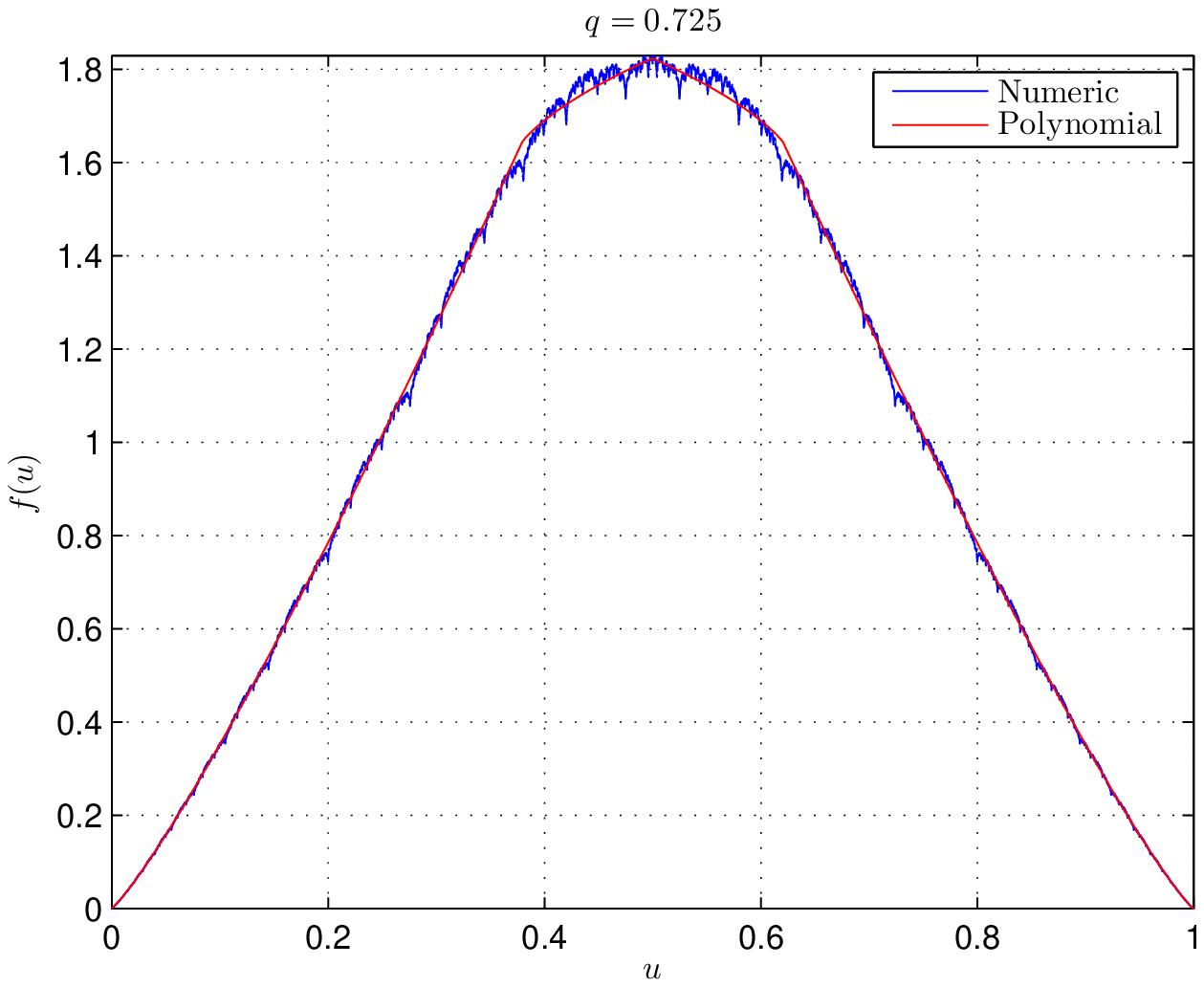}}\subfigure[]{\includegraphics[width=.5\linewidth]{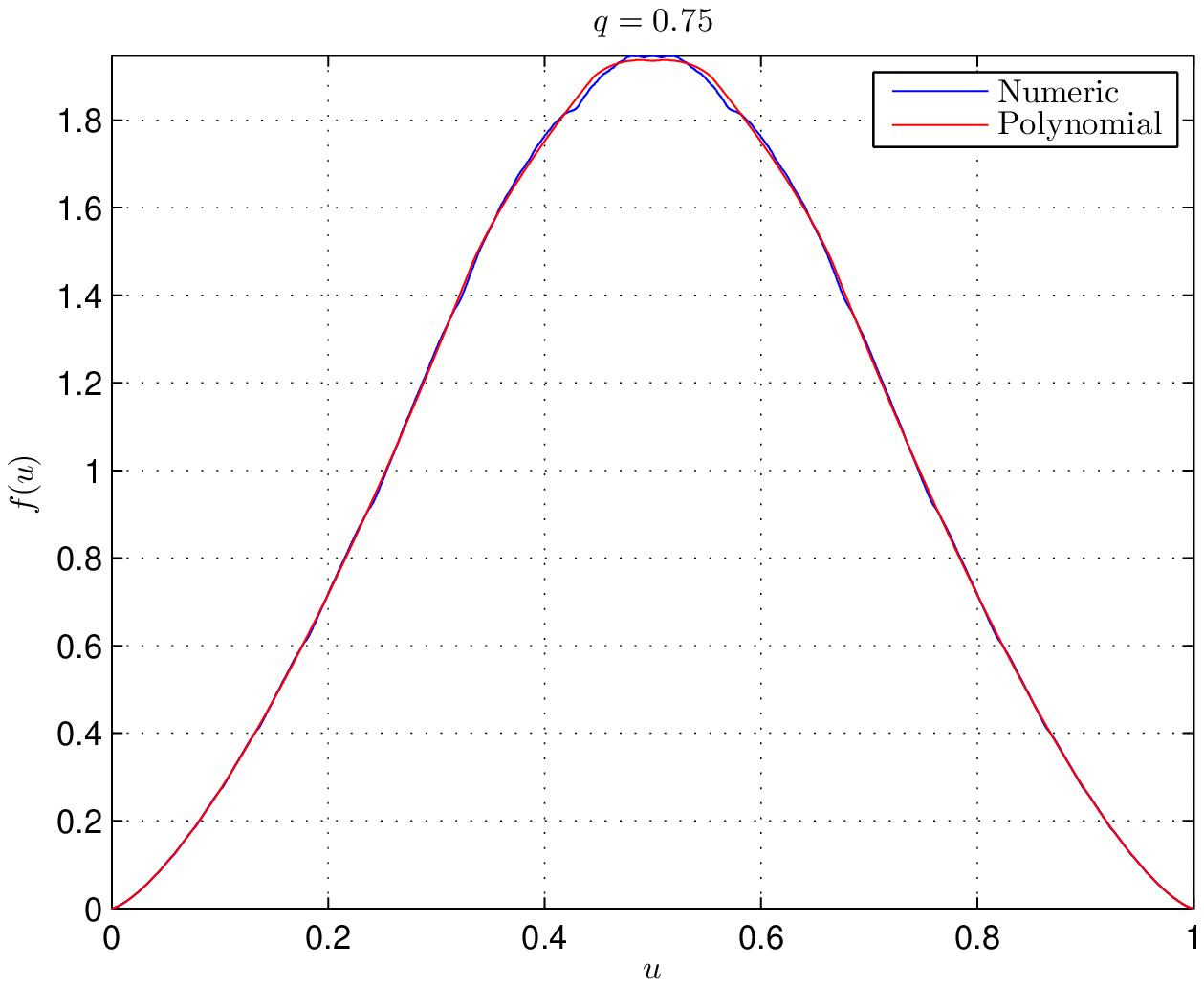}}
\subfigure[]{\includegraphics[width=.5\linewidth]{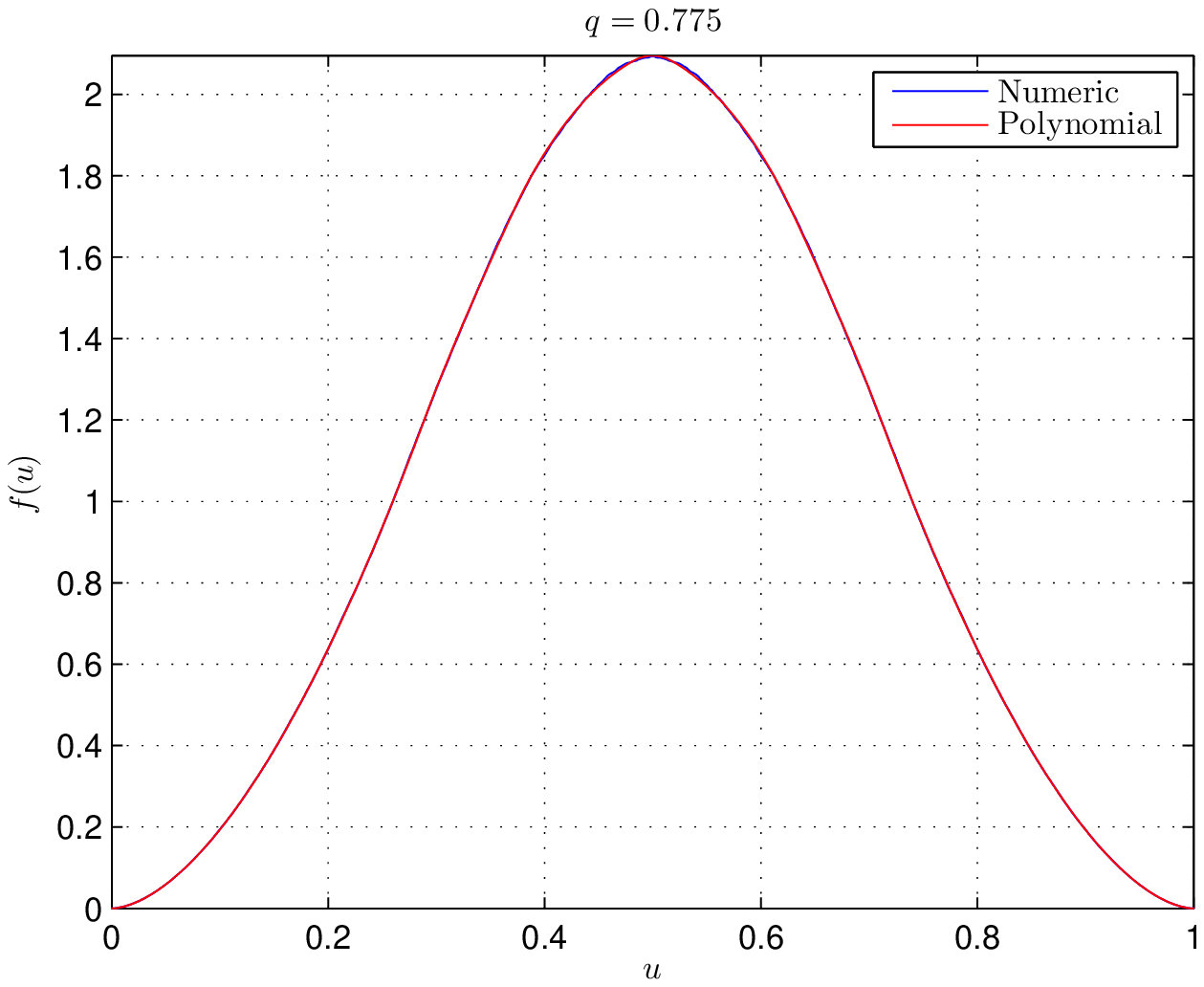}}\subfigure[]{\includegraphics[width=.5\linewidth]{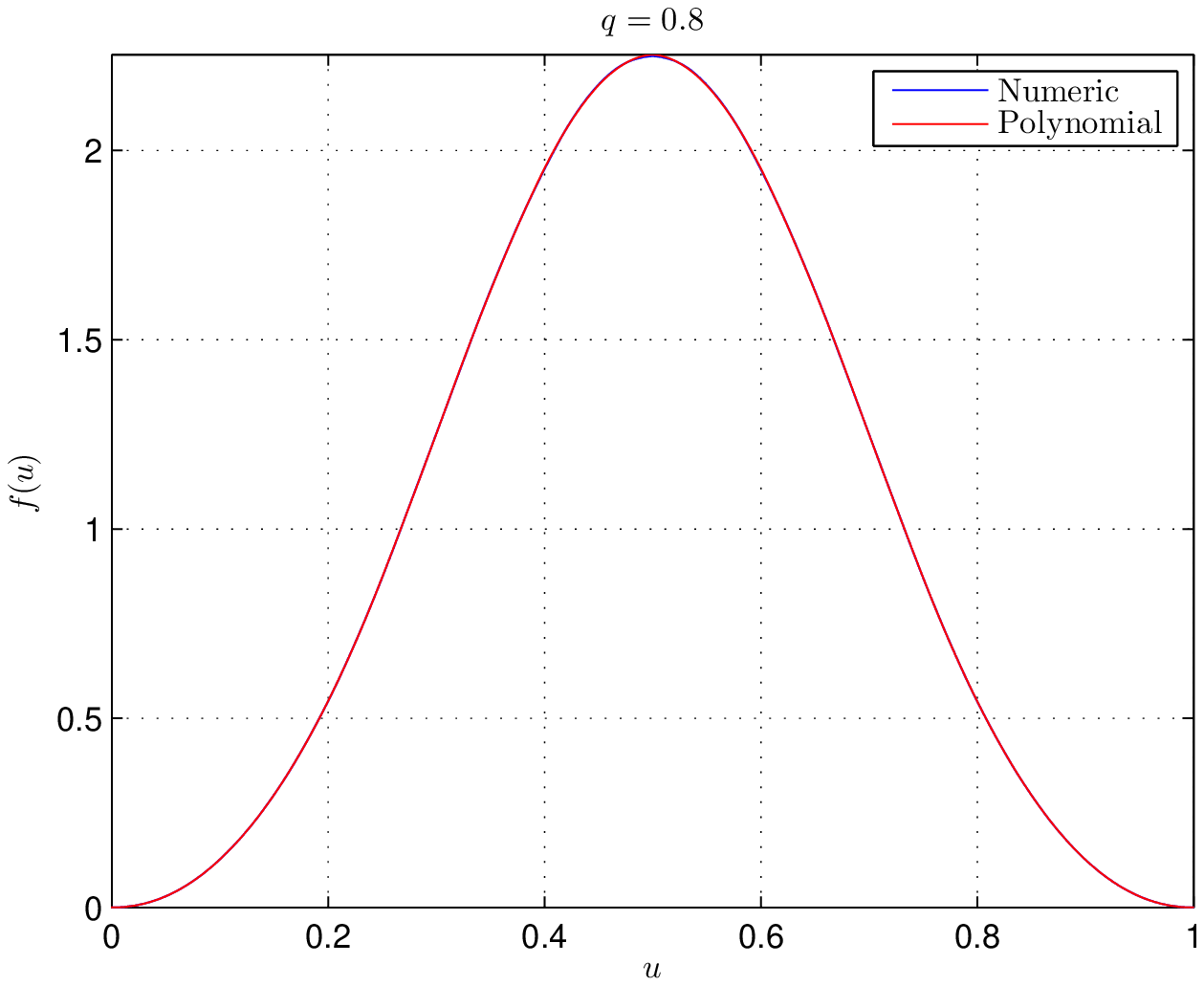}}
\caption{Comparisons of polynomial approximation with numeric approximation, where $N=10^5$ and $\delta = 10^{-10}$ for numeric approximation. These results show that polynomial approximation fits numeric approximation very well. Especially, as $q$ increases, polynomial approximation almost coincides with numeric approximation. (a) $q$ = 0.725. (b) $q$ = 0.75. (c) $q$ = 0.775. (d) $q$ = 0.8.}
\label{fig:polynomial}
\end{figure*}

Some examples of polynomial approximation have been included in Fig. \ref{fig:polynomial}. Considering the complexity, only the results for $1/\sqrt{2} \leq q \leq 0.8$ are reported. Fig. \ref{fig:polynomial} shows that in general, the curves of polynomial approximation fit those of numeric approximation very well. Especially, as $q$ increases, the curves of polynomial approximation almost coincide with those of numeric approximation. In addition, Fig. \ref{fig:polynomial}(a) also shows the affirmation in \cite{FangSPL09} may fail because $q > 1/\sqrt{2}$ does not guarantee smooth $f(u)$. Nevertheless, $f(u)$ does become less irregular as $q$ increases.

\section{Gaussian Approximation at Very Low Rates}\label{sec:gaussian}
As pointed out in Section \ref{sec:polynomial} that as $q$ increases, polynomial approximation to $f(u)$ becomes very complex. Thus a simpler approximation method is needed at very low rates. Through experiments, we observe that $f(u)$ becomes bell-shaped at very low rates \cite{FangSPL09}. This phenomenon suggests that a Gaussian function centered at 0.5 may be good approximation to $f(u)$, i.e. 
\begin{equation}
f(u) \approx \frac{1}{\sqrt{2\pi}\sigma}\exp\left(-\frac{(u-0.5)^2}{2\sigma^2}\right).
\end{equation}
Obviously, the problem now boils down to how to estimate $\sigma^2$ for given $q$.

\subsection{Estimation of $\sigma^2$}
Here we propose a simple method to estimate $\sigma^2$ by exploiting $qf(0.5) = f(0.5/q)$ [Fig. \ref{fig:dacdis}]. For a large $q$, we have
\begin{equation}
qf(0.5) \approx \frac{q}{\sqrt{2\pi}\sigma}
\end{equation}
and
\begin{equation}
f(0.5/q) \approx \frac{1}{\sqrt{2\pi}\sigma}\exp\left(-\frac{(1-q)^2}{8q^2\sigma^2}\right).
\end{equation}
Hence,
\begin{equation}
q \approx \exp\left(-\frac{(1-q)^2}{8q^2\sigma^2}\right).
\end{equation}
Therefore
\begin{equation}
\sigma^2 \approx -\frac{(1-q)^2}{8q^2\ln{q}}.
\end{equation}

\subsection{Simulation Results}
\begin{figure*}
\centering
\subfigure[]{\includegraphics[width=.5\linewidth]{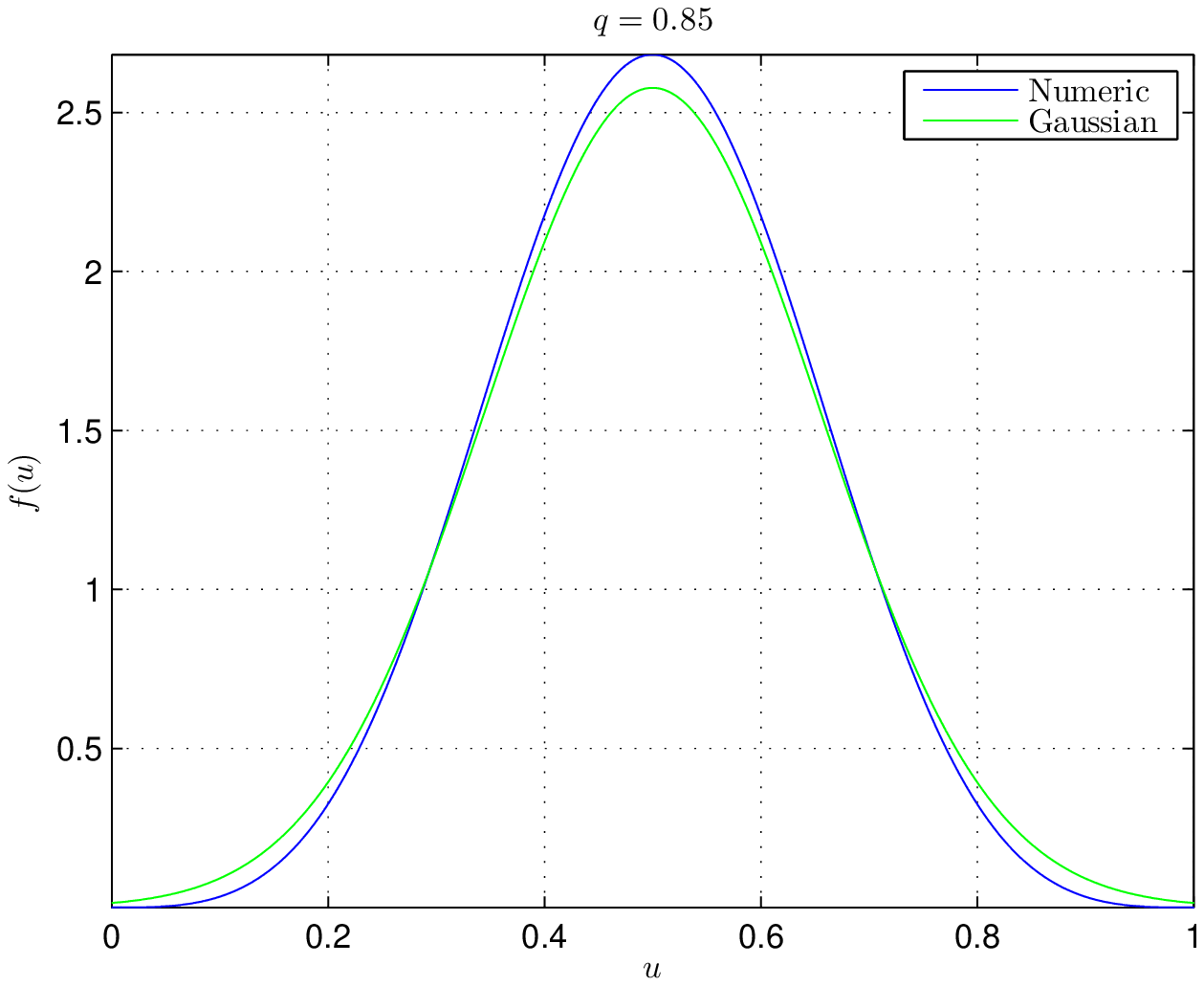}}\subfigure[]{\includegraphics[width=.5\linewidth]{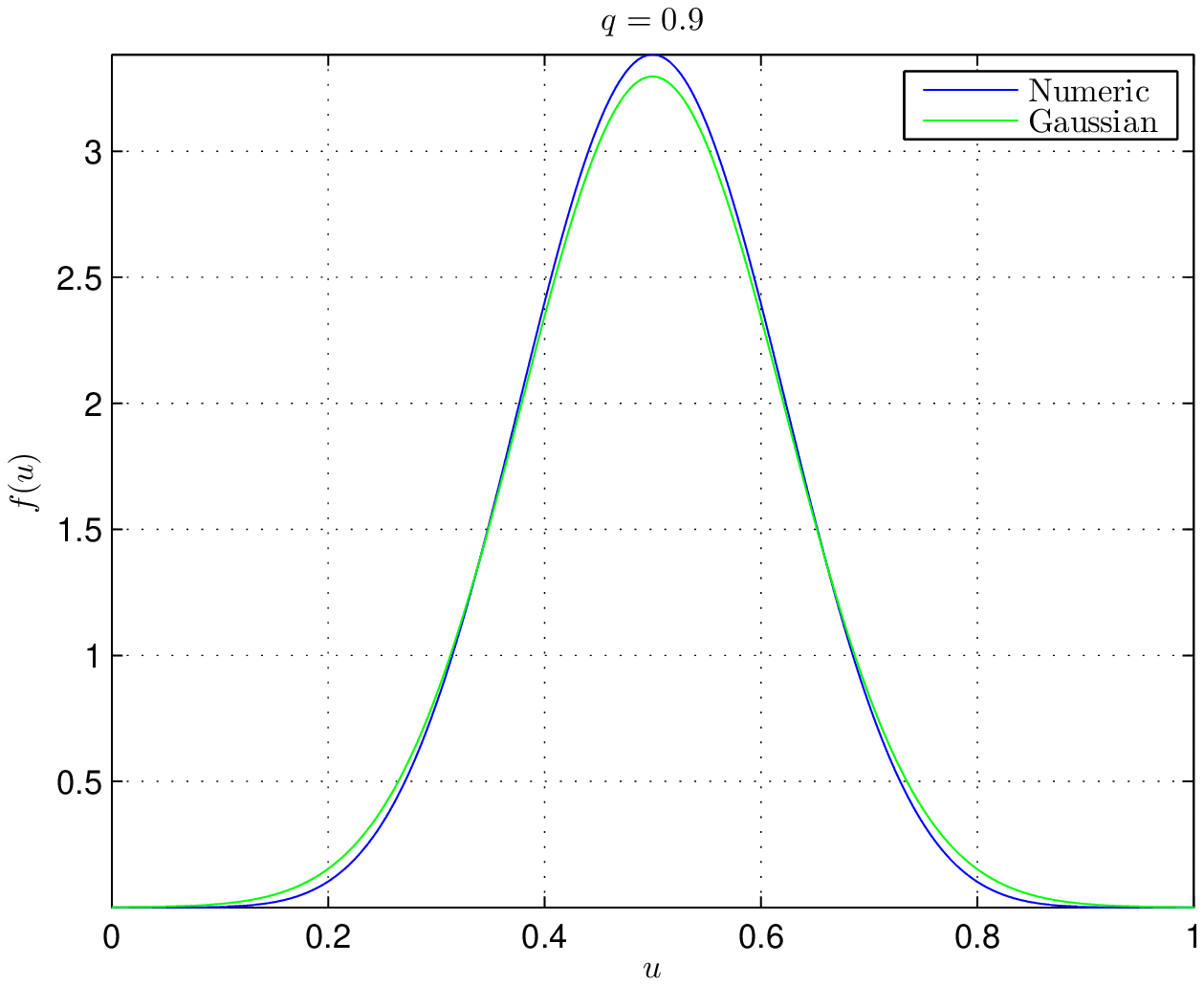}}
\subfigure[]{\includegraphics[width=.5\linewidth]{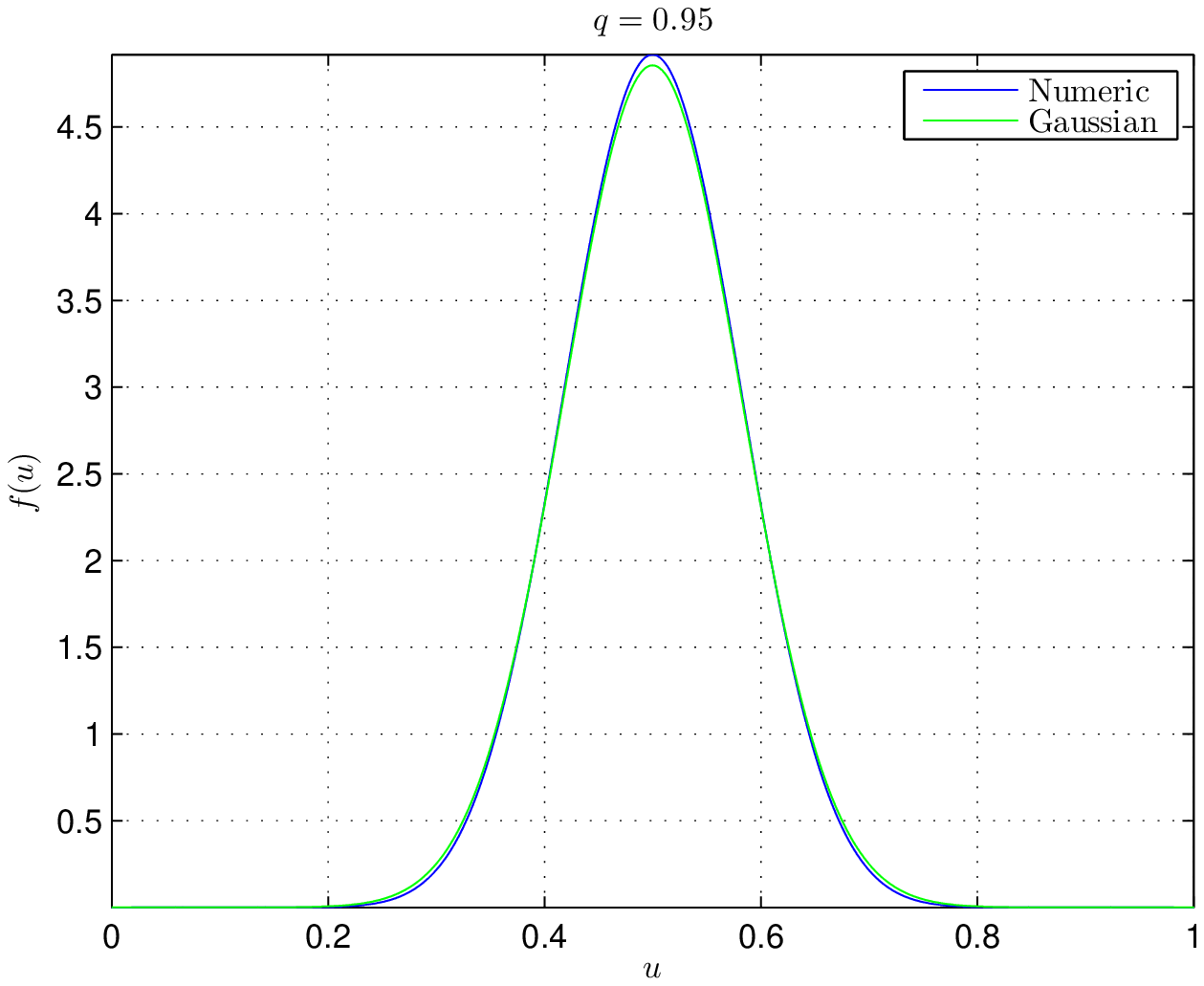}}\subfigure[]{\includegraphics[width=.5\linewidth]{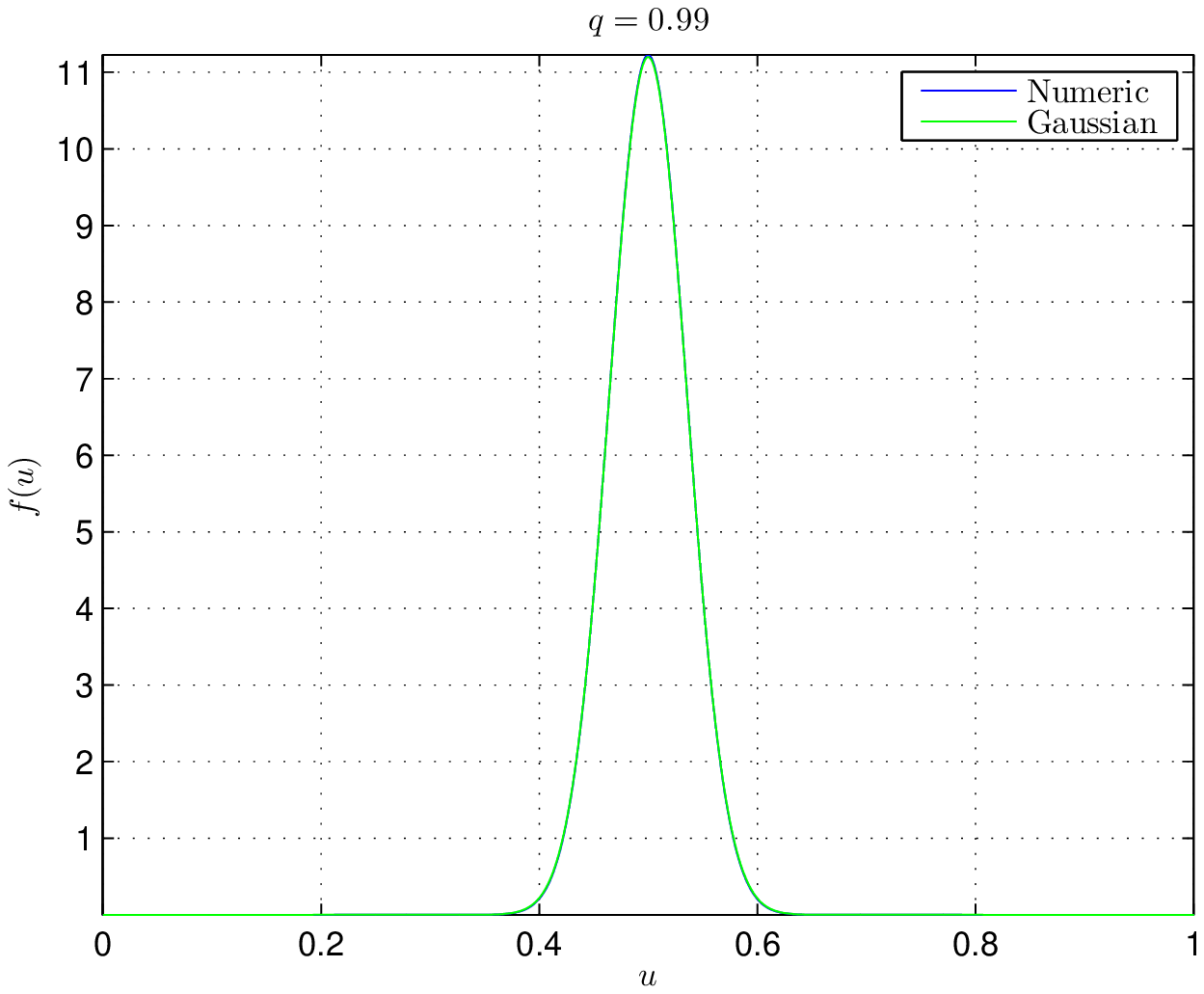}}
\caption{Comparisons of Gaussian approximation with numeric approximation, where $N = 10^5$ for numeric approximation. As $q$ increases, Gaussian approximation becomes more and more accurate. (a) $q = 0.85$. $\delta = 10^{-10}$ for numeric approximation. (b) $q = 0.9$. $\delta = 10^{-10}$ for numeric approximation. (c) $q = 0.95$. $\delta = 10^{-10}$ for numeric approximation (d) $q = 0.99$. $\delta = 10^{-9}$ for numeric approximation.}
\label{fig:gaussian}
\end{figure*}
 
Some examples of Gaussian approximation are included in Fig. \ref{fig:gaussian}. These plots show that as $q$ increases, the curves of Gaussian approximation become closer and closer to those of numeric approximation. Especially, when $q=0.99$, the curve of Gaussian approximation almost coincides with that of numeric approximation. All these results confirm that Gaussian approximation does work well at very low rates.

\section{Conclusion}\label{sec:conclusion}
This paper proposes three approximation methods for DAC codeword distribution of equiprobable binary sources along proper decoding paths. These methods are well justified by simulation results. The related software is available on \cite{FangHomepage}. 

Nevertheless, there remain many open issues. Firstly, how to format the problem for codeword distribution along wrong decoding path? Secondly, for general (non-equiprobable or $M$-ary) sources, how to format the problem? Thirdly, can we find the number of possible decoding paths as well as the distributions of $D(X, \tilde{X})$ and $D(Y, \tilde{X})$, for a given DAC code of $X$. Finally, it is an interesting issue to define codeword distribution for the ECAC.


\end{document}